\documentclass[12pt,preprint,natbib]{aastex}
\renewcommand{\baselinestretch}{1.5}

\RequirePackage{lineno} 
\usepackage{epstopdf}  

\begin{document}
\bibliographystyle{apj}

\begin{singlespace}

\setlength{\footskip}{0pt} 

\title{Will Comet ISON (C/2012 S1) Survive Perihelion?}

\author{Matthew M. Knight\altaffilmark{1,2,3}, Kevin J. Walsh\altaffilmark{4}}

\author{Submitted to the {\it Astrophysical Journal Letters}}
\author{Received: 2013 Aug 13; Revised: 2013 Sep 3; Accepted: 2013 Sep 7}

\altaffiltext{1}{Contacting author: knight@lowell.edu.}
\altaffiltext{2}{Lowell Observatory, 1400 W. Mars Hill Rd, Flagstaff, AZ 86001, USA}
\altaffiltext{3}{Visiting scientist at The Johns Hopkins University Applied Physics Laboratory, 11100 Johns Hopkins Road, Laurel, Maryland 20723, USA}
\altaffiltext{4}{Southwest Research Institute, 1050 Walnut St., Suite 400, Boulder, CO 80302, USA}

\begin{abstract}
  On 2013 November 28 Comet ISON (C/2012 S1) will pass by the Sun with
  a perihelion distance of 2.7 solar radii. Understanding the possible
  outcomes for the comet's response to such a close passage by the Sun
  is important for planning observational campaigns and for inferring
  ISON's physical properties.  We present new numerical simulations
  and interpret them in context with the historical track record of
  comet disruptions and of sungrazing comet behavior. Historical data 
  suggest that sizes below $\sim$200~m are susceptible to destruction 
  by sublimation driven mass loss, while we find that for ISON's 
  perihelion distance, densities lower than 0.1~g~cm$^{-3}$ are 
  required to tidally disrupt a retrograde or non-spinning body. Such low
  densities are substantially below the range of the best-determined
  comet nucleus densities, though dynamically new comets such as ISON
  have few measurements of physical properties.  Disruption may occur
  for prograde rotation at densities up to 0.7~g~cm$^{-3}$, with the
  chances of disruption increasing for lower density, faster prograde
  rotation, and increasing elongation of the nucleus.
Given current constraints on ISON's nucleus properties and the typically 
determined values for these properties among all comets, we find tidal 
disruption to be unlikely unless other factors (e.g., spin-up via
torquing) affect ISON substantially. Whether or not disruption occurs,
the largest remnant must be big enough to survive subsequent mass loss 
due to sublimation in order for ISON to remain a viable comet well after 
perihelion.
\end{abstract}


\section{INTRODUCTION}
\label{sec:intro}
Comet ISON (C/2012 S1) was discovered on 2012 September 21 and will reach perihelion on 2013 November 28 at a sungrazing distance of 0.0125 $\mathrm{AU}=\mathrm{2.7}$ solar radii ($R_\odot$; \citealt{cbet3238}). The lead time of more than one year from discovery until perihelion is unique in sungrazing comet history and allows unprecedented planning of ground- and space-based resources in order to maximize the scientific return. While ISON's orbit is well known, the likely result of its perihelion passage is not, prompting this $Letter$. 

Understanding the possible outcomes for the comet's response to such a close passage by the Sun is important for a variety of reasons. First, ISON is much better placed for Earth-based observations post-perihelion than pre-perihelion, 
 so most intensive observational campaigns are planned for after perihelion. If ISON is unlikely to survive then all efforts should be made to obtain data prior to perihelion. Second, ISON is expected to be observable by the fleet of space-based solar observatories, most of whom have limited real-time flexibility in their observations; early predictions of ISON's behavior will help them optimize their observing sequences. Third, it has recently become possible to use comets as probes of the solar environment \citep{schrijver12,bryans12,downs13}. These studies require basic assumptions about the physical nature of the comet; in order to properly interpret these data it is critical to understand whether ISON behaves differently than the two previous comets utilized in this manner.

As detailed in the remainder of Section~\ref{sec:intro}, there are numerous studies of tidal encounters of comets and asteroids with planetary bodies, and a historical track record of sungrazing comets being destroyed by disintegration and/or disruption. We use these as a guide and conduct numerical simulations of ISON's perihelion encounter in Section~\ref{sec:sims}. Finally, in Section~\ref{sec:disc} we interpret these simulations and make reasonable assumptions about ISON's nuclear properties in order to determine the range of likely outcomes and the comet's chances of surviving perihelion.

\subsection{Tidal Disruption}
Disruptions of solar system bodies due to tidal forces can be
spectacular, with Comet Shoemaker-Levy 9 (henceforth SL9)
a prime example. While \citet{roche47} theorized about the disruption
of fluid bodies on circular orbits with synchronized spin states, SL9
and other recent discoveries have motivated more sophisticated studies
of different types of encounters and more realistic models of internal 
structure.

Oort cloud comets typically have nearly parabolic encounters with the
Sun, which is a vastly different geometry than that envisioned
by Roche
(see also \citealt{chandrasekhar69}). Similarly, a
large body of work based on decades of observations and numerical
simulations has painted a picture of small Solar System bodies as
``rubble piles,'' again vastly different than the liquids considered
by Roche or Chandresekhar (see \citealt{richardson02} for a
review). These gravitationally bound bodies are typically assumed to
have zero or possibly very limited tensile strength, and therefore may
rely only on their self-gravity to bind them and the shear strength
afforded by their building blocks' physical sizes to resist re-shaping.
In fact, it was the breakup of comet Ikeya-Seki in 1965
(\citealt{sekanina66} and references therein) during its sungrazing passage 
that prompted \citet{opik66} to suggest that it may have been a 
``heap of rubble.''

Prior to the discovery of SL9, work was underway to extend the
calculations of Roche to a more diverse set of
circumstances. \citet{boss91} used smoothed particle hydrodynamics
simulations to model inviscid planetesimals passing close to
Earth, finding that 
the increase of the
planetesimal's spin angular momentum during the encounter can induce
equatorial mass-shedding. \citet{sridhar92} extended the analytical
work into the regime of viscous-fluid bodies having
parabolic encounters with a planet, finding that mass shedding begins 
at $\sim$69\% of the classical Roche limit in this scenario.

The discovery of SL9 in 1993 launched significant numerical modeling
efforts of tidal disruption outcomes. Specifically, $N$-body
gravitational codes were employed to model gravitationally bound
``rubble pile'' bodies' constituent pieces.  In seeking a match to the
precise morphology observed in the SL9 fragment
chain, \citet{asphaug96} identified many important and challenging
degeneracies in the tidal breakup process. While the ``strength'' of
the breakup increased with decreasing density and increasing prograde
spin rate, the trade off between density and spin rate was very clean 
-- there is no simple way to break that 
degeneracy. \citet{richardson98} went further and explored the effects
of shape during a tidal encounter,
finding that elongated bodies can disrupt even more violently, but in
a manner that is very dependent on the orientation of their long axis
at the time of close approach. Recent work for the SL9 disruption has
found that using non-spherical particles for the rubble pile
progenitor frustrates disruption somewhat \citep{movshovitz12},
suggesting that hard spherical particle models (as presented here) may
represent an upper limit for disruption parameters.

However, there are numerous examples of cometary breakups that do not
fit neatly into any of the previous ``rubble pile'' models, including
nearly all of those that suffered disruption far from perihelion or
any other known perturbations (e.g., \citealt{boehnhardt04}). 
These events remind us that while
models might predict only a modest re-shaping or very weak tidal forces
acting across a body, it might only need to ``light a fuse'' inside a
comet leading to a later disruption.

There are substantial uncertainties in modeling a cometary disruption
near the Sun. Specifically, the extensive modeling of the SL9 event
presented in \citet{asphaug96} inferred very low or zero tensile
strength to best explain the observations, but 
\citet{holsapple08} demonstrated that even very small amounts of
tensile strength can have a dramatic effect on disruption limits,
especially at $\sim$km sizes. While tensile strength is very
hard to measure, it was inferred for 9P/Tempel 1 from the Deep Impact
experiment as being considerably weaker than heavily fractured ice,
but non-zero \citep{richardson07,holsapple07}. However, given the
extremely high evaporation rates during perihelion,
there are other forces that may exceed any small values of
tensile strength. An example demonstrated by \citet{gundlach12} finds
that 
the forces created during the outgassing of ices on a sungrazing comet
may be strong enough that the reaction force on the nucleus delays or
prevents tidal splitting.

Complications aside, for ISON's perihelion distance, the classical
Roche limit predicts mass loss for densities $<$1.03~g~cm$^{-3}$,
while \citet{sridhar92} predicts mass loss for densities
$<$0.34~g~cm$^{-3}$.  \citet{holsapple08} predicts
mass loss for densities $<$0.13~g~cm$^{-3}$ for a non-spinning
spherical cohesionless rubble pile with a 40$^\circ$ angle of
friction, while a similar body with a lower angle of friction
(20$^\circ$) would only need a density $<$0.27~g~cm$^{-3}$. Though
``disruption'' is defined differently in each work, all
estimated comet
densities have been $<$1.0 g cm$^{-3}$ \citep{weissman04,ahearn11b}, 
so ISON clearly warrants further investigation.

Historically, the term ``sungrazing'' was applied exclusively to
members of the Kreutz family (\citealt{kreutz88}; discussed below). 
When {\it Solar and Heliospheric Observatory (SOHO)} began discovering non-Kreutz
comets in small-perihelion distance ($q$) orbits, ``sungrazing'' and
``sunskirting'' began to be used to distinguish the Kreutz from the
larger-$q$ near-Sun comets, but without being formally defined. Rather
than dividing at some arbitrary distance, we propose to give it a
physical basis by using the Roche limit, which can be calculated
easily if one measures the comet's density or can safely assume
a nominal value. This limit distinguishes between comets that may suffer
tidally driven mass loss (sungrazers) from those that will not
(sunskirters). With this definition, the only ground-observed comets 
classified as sungrazers that are not Kreutz comets are C/1680 V1 and ISON
\citep{marsden08}.

\subsection{Previous Disruptions of Sungrazing Comets}
Observational constraints on the behavior of sungrazing comets near perihelion come entirely from the well known Kreutz group. Kreutz comets have $q=1$--2 $R_\odot$
and orbital periods of 500--1000 yr. They are dynamically linked to a single progenitor, with the members having been produced by cascading fragmentation over several orbits (cf.\ \citealt{marsden67,marsden89,sekaninachodas07}). The group consists of a smattering of naked-eye comets seen over several centuries plus a nearly continuous stream of faint comets observed only with space-based coronagraphs. 

The coronagraphically discovered Kreutz comets do not survive perihelion; in fact, they typically peak in brightness at 10--15 $R_\odot$ then rapidly fade \citep{biesecker02,knight10d}. Four Kreutz comets -- C/1880 C1, C/1887 B1, C/1945 X1, and C/2011 W3 (Lovejoy) -- survived until or shortly after perihelion, often appearing as headless tails receding from the Sun, while five -- C/1843 D1, C/1882 R1, C/1963 R1, C/1965 S1 (Ikeya-Seki), and C/1970 K1 -- clearly survived \citep{sekanina02b,kronk03,kronk09,kronk10,sekanina12}. Other than C/1970 K1, all that clearly survived were observed to split during the perihelion passage, with at least one substantial remnant surviving and exhibiting cometary behavior for an extended time.

Thus, the Kreutz comets demonstrate the range of outcomes that may
befall ISON. The smallest comets, consistently estimated to be 
$<$0.2 km in radius (e.g.,
\citealt{macqueen91,raymond98,sekanina03,knight10d}) succumb to
sublimation driven mass loss prior to perihelion. Intermediate sized
comets, 0.2--1.0 km in radius, are large enough to survive mass loss
due to sublimation \citep{iseli02,sekanina03}, but likely disrupt with
no individual fragment sufficiently large to remain a viable comet
significantly beyond perihelion. The largest comets (radius $>$1 km)
easily survive mass loss due to sublimation alone and, even if they 
fragment, remain viable
comets that will return on a subsequent perihelion passage.

\subsection{Comet ISON}
\label{sec:introison}
Despite its sungrazing orbit, ISON is not a member of the Kreutz group; it is a ``dynamically new'' comet entering the solar system for the first time (as evidenced by its reciprocal original semi-major axis of 7$\times$10$^{-6}$ AU$^{-1}$),
whereas the Kreutz comets have made at least several sungrazing orbits over the last few thousand years. There are expected to be differences in the outer layers of dynamically new comets as compared to returning comets due to the former's long residence in the Oort cloud (cf. \citealt{stern90}). While we use the Kreutz comets as a guide out of necessity, this differing evolutionary history may result in compositional or structural differences that affect ISON's survivability. 

As  discussed in the preceding subsection, the first criterion for ISON to survive perihelion is its nuclear size. Hubble Space Telescope (HST) observations set an upper limit on the radius of $\sim$2 km \citep{cbet3496}. We estimate the minimum radius to be $\sim$0.4 km using the only published gas measurements \citep{iauc9254,iauc9257} and following the methodology of \citet{cowan79} (using M. A'Hearn's web-based calculator\footnote{http://www.astro.umd.edu/{\raise.17ex\hbox{$\scriptstyle\sim$}}ma/evap/index.shtml}). Similar calculations using the estimated production rates of CO or CO$_2$ from Spitzer Space Telescope images \citep{cbet3598} and upper limits of CO from HST spectroscopy (M. A'Hearn, private comm.) yield minimum radius estimates of 0.1--0.3 km. 

This range of possible nuclear sizes suggests that, based on the behavior of Kreutz comets, 
ISON is likely large enough to survive mass loss due to sublimation alone, but not necessarily so large that some fragment(s) would reasonably be expected to remain viable if the nucleus disrupts. Note that ISON's perihelion distance is slightly larger than that of the Kreutz family so it will likely suffer less sublimation driven mass loss; however, for simplicity we assume it has the same survival thresholds. Therefore, in order to estimate ISON's chances of survival, we need to investigate its susceptibility to tidal disruption.

\section{SIMULATIONS}
\label{sec:sims}
The history of cometary disruption and fragmentation is strongly
suggestive of forces beyond gravity playing a significant
role (e.g., C/1999 S4 LINEAR, 73P/Schwassman-Wachmann 3).
However, we aim to build a baseline for the expected behavior of
ISON during its perihelion passage strictly due to the tidal
forces. From this baseline, with historical perspective and the best
possible estimates of important unknown variables, we make a
prediction about its behavior at, and immediately following,
perihelion.

\citet{richardson98} explored a range of
shape and spin combinations and a wide range of close encounter
distances and hyperbolic encounter characteristics for close passages 
of rubble-pile asteroids to the Earth. Here, the
encounter parameters are known with significant accuracy, and the
encounter is nearly parabolic.
Given that this specific encounter is of great interest, and velocity at 
infinity $(v_\infty)=0$~km~s$^{-1}$ (e.g., a comet arriving from the 
Oort cloud) was not explicitly tested in \citet{richardson98}, we 
endeavor to explore more deeply the possible outcomes for 
$q=2.7$ $R_\odot$ and $v_\infty\sim0$~km~s$^{-1}$, focusing on 
the effects of internal density, spin, and shape of the body.

Our simulations  were designed to test a parabolic
flyby of a gravitational aggregate, and used the $N$-body code {\tt
pkdgrav}
\citep{richardson00}. The progenitor bodies were constructed
of $\sim$2000 hard spherical particles that were initially in a
close-packed configuration. The encounters were started at 10 $R_\odot$
and continued until the re-accumulation of fragments was considered to
be complete -- typically 10,000--20,000 timesteps, each of
$\sim$50~s. Starting the simulations earlier or running them later 
has no effect on the results \citep{walsh06}.
As will be described below, bulk densities were explored
in the range of 0.075--0.8~g~cm$^{-3}$, and two shapes were explored, a
sphere and a body with 2:1:1 axis ratios rotating uniformly around its short
axis (the elongated body is near the extreme axis ratios observed for
comets and should represent an end-member case of all the possible shapes).

The rotation rate selected for the prograde encounters was taken to be
50\% of the critical rotation rate (the rate at which mass begins to 
be lost due to centrifugal acceleration), 
following the simplistic formulation
\begin{equation}
P_\mathrm{crit}=\frac{3.3~\mathrm{hr}}{\sqrt{\rho}}\sqrt{(a/b)}
\end{equation}
where $P_\mathrm{crit}$ is the rotation period corresponding to the
critical rotation rate, $\rho$ is the density of the body in
g~cm$^{-3}$, and $a$ and $b$ are the long and intermediate axis
lengths respectively (using the same notation as in
\citealt{richardson98}). While the selection of this rotation rate was
somewhat arbitrary, much more rapid rotation would dramatically
increase the chances of disruption and such fast rotation rates are
rare among comets.  Thus for a spherical body ($a=b$) with
$\rho=0.5$ g~cm$^{-3}$, $P_\mathrm{crit\_50\%}=9.33$~hr. 
For the same densities but an elongated body
($a=2\times b$) the rotation periods were a factor of $\sqrt{2}$
longer.

For the baseline simulations with non-rotating (which is akin to a
very long rotation period in these simulations) or prograde rotating
spherical bodies, 10 simulations for each density were run to account
for any geometry inherent in the close-packed configuration of the
progenitor rubble pile. For the cases with an elongated body, we
tested 30 different encounters because of the substantial changes in
outcome depending on the alignment of the body's long axis at
perihelion. Disruption is more likely if the spin angular momentum is
aligned with the encounter angular momentum \citep{richardson98},
though as the spin rate decreases this becomes less important. 
  Elongation is less important in the non-spinning case and so only
  spherical bodies were tested. No cases of retrograde rotators were
tested as they are less likely to disrupt than non-spinning cases 
\citep{richardson98}, and the results below show non-spinning as a
very unlikely disruption scenario.

\begin{figure}
  \centering
  \epsscale{0.7}
  \plotone{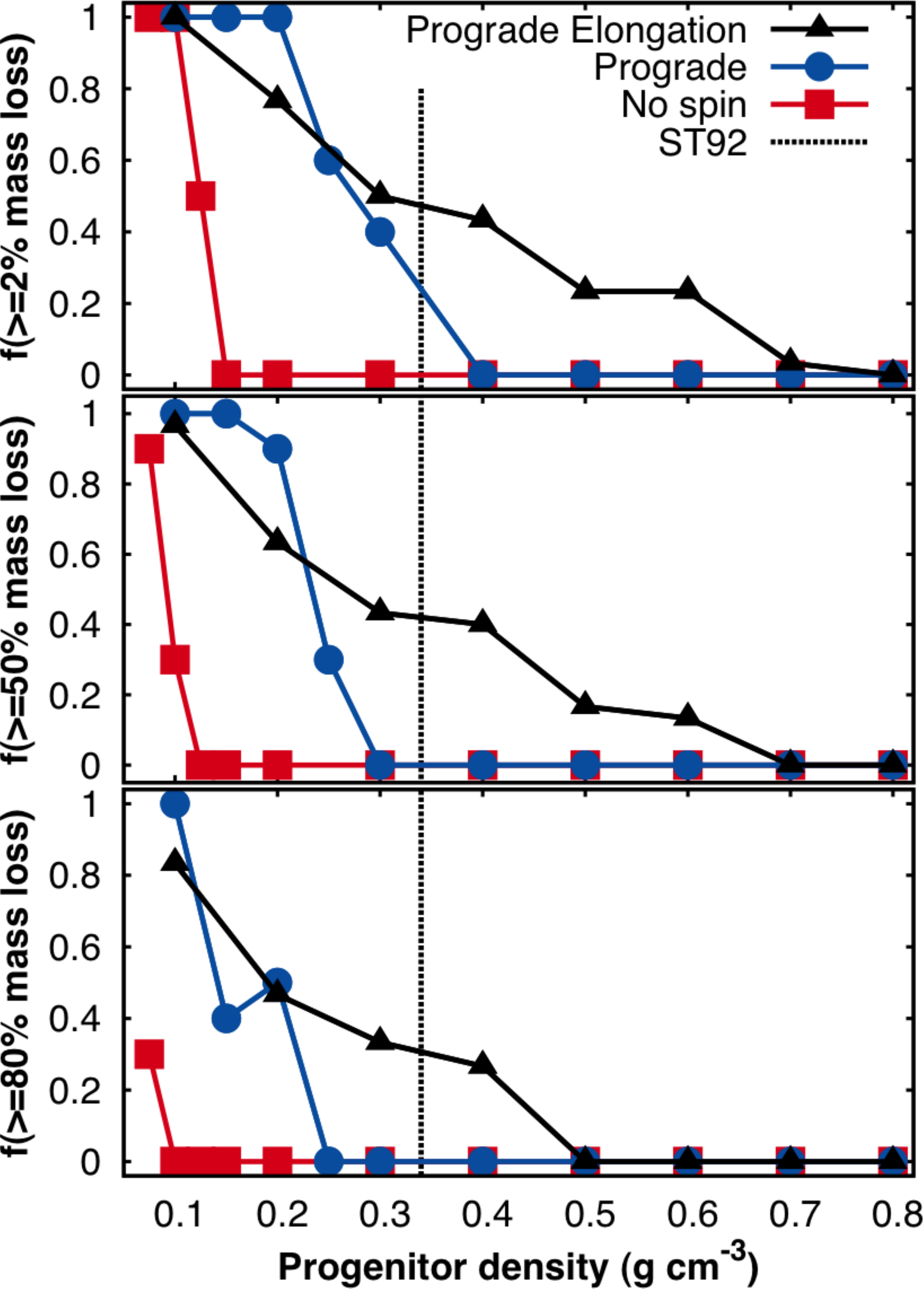}  
  \caption[Fraction of simulations with mass loss above various thresholds]{Fraction of simulations with mass loss exceeding 2\% (top), 50\% (middle), and 80\% (bottom) as a function of density. Three cases are plotted: non-spinning spherical nucleus (red squares), prograde spherical nucleus (blue circles), and prograde elongated nucleus (black triangles). No retrograde cases were plotted because mass loss is less likely than for the non-spinning case and therefore only occurs for extremely low densities ($<$0.1~g~cm$^{-3}$). The vertical dashed line shows the density at which \citet{sridhar92} predict ISON should begin to shed mass.
}
  \label{fig:mlr}
  \label{lastfig}
\end{figure}

The results of these simulations are presented in Figure~\ref{fig:mlr}, 
which shows the fraction, $f$,  of simulations in which mass loss 
exceeded 2\%, 50\%, and 80\% of the total mass
of the progenitor for each of the tested bulk density values. 
A conservative estimate of survival is given by $1-f_\mathrm{2\%}$, while $f_\mathrm{80\%}$ depicts catastrophic disruptions.
For bodies with initially no rotation, there is a sharp increase in
the degree of the disruption as seen by the increasing $f$ for
densities $<$0.15~g~cm$^{-3}$. The disruptions begin at higher
densities for the bodies with prograde rotation, with disruptions
beginning at 0.3~g~cm$^{-3}$ in the spherical case. Elongated bodies
with prograde rotation, the extreme case tested, experience mass loss
even earlier, with some bodies at densities up to 0.7~g~cm$^{-3}$
disrupting. The cases with elongation had nearly bi-modal results,
with either no mass-loss or dramatic mass loss due to the importance
of the long-axis location at perihelion \citep{richardson98}.
Disruption did not occur in any simulations in which the density was
$>$0.7~g~cm$^{-3}$ .

\section{DISCUSSION}
\label{sec:disc}
The first three columns of Table~\ref{t:parameters} give the nuclear
parameters that affect survival, their considered ranges, and a
qualitative description of their effect on survivability. We
specifically tested a range of density, axis ratio, and sense of
rotation. The simulations are scale invariant, and thus can be
translated to any pre-encounter nuclear size, so in effect radius has
also been sampled (noting that the largest remnant needs to be
$\gtrsim$0.1 km at perihelion to survive the remainder of the
apparition). Therefore, the only parameter not specifically tested is
rotation period, which has a well understood degeneracy with density
\citep{asphaug96,richardson98}. Rotation periods potentially span a
tremendous range that, if fully explored, would have been too
CPU-intensive for the current $Letter$. The effect of rotation period
can, however, be visualized in Figure~\ref{fig:mlr} by 
shifting the curves up and to the right for shorter prograde
rotation periods (mass loss due to tidal forces
results at larger densities) or down and to the left for longer prograde, 
or any retrograde, rotation periods (mass loss due to tidal forces only
occurs at very low densities).

The last column in Table~\ref{t:parameters} gives the likely range of 
values for each parameter for ISON. While radius (discussed in 
\S\ref{sec:introison}) and 
sense of rotation (retrograde; 
provided by T. Farnham based on the pole orientation from \citealt{cbet3496})
are partially constrained by published observations, the rest are not. 
Comet nuclear properties are notoriously difficult to measure, and the 
properties of dynamically new comets like ISON are almost completely 
unconstrained. Therefore, the ranges of the remaining parameters cover 
the known values of all comets as compiled from \citet{lamy04}, 
\citet{samarasinha04}, \citet{weissman04}, \citet{ahearn11b}, and our 
own literature search for more recent publications.

ISON appears likely to survive the combination of mass loss 
due to sublimation and tidal disruption for most plausible scenarios. 
If it is ``typical'' -- radius $\sim$1 km, density $\sim$0.5~g~cm$^{-3}$, 
axis ratio $\sim$1, rotation period $\sim$24 hr, and a random sense of
rotation (or retrograde as the preliminary results suggest) 
-- it is very likely to survive the encounter. 
Given that comet densities are
relatively well constrained to be near $\sim$0.5~g~cm$^{-3}$, the
rotation period is the parameter whose plausible range poses the
largest threat to ISON's ability to survive perihelion. For a density
of $\sim$0.5~g~cm$^{-3}$, ISON would lose mass for many scenarios in
which the rotation period was prograde and faster than $\sim$9 hr,
with tidally driven mass loss increasing as rotation period decreases. 
Roughly 30\% of
measured comet rotation periods are $<$9 hr, although this is likely 
overestimated because shorter rotation periods are easier to measure.
Furthermore, assuming that sense of rotation is random, half of these
fast rotators would be retrograde and therefore unlikely to disrupt.

As of 2013 August, none of the parameters in our simulations have been tightly constrained for ISON, although some (most likely rotation period and/or pole orientation) may yet be determined prior to perihelion. 
However, even if these quantities are constrained, \citet{samarasinha13} showed that ISON's rotation is likely to be highly excited near perihelion so values determined at larger heliocentric distances may not hold through perihelion, and mass loss due to spin-up may occur. 
Further investigations of this possibility are beyond the scope of this $Letter$. 
Also beyond the scope are the effects of sublimation driven mass loss during the encounter, although we do not expect it to alter our results substantially.

We hope that observations around and after perihelion may yield enough information to infer many of these parameters and therefore conclusively determine ISON's susceptibility (or lack thereof) to tidal disruption. Furthermore, the current work should serve as a guide for studies after perihelion to estimate ISON's density based on measurable quantities and the results of the perihelion passage.

\section*{ACKNOWLEDGMENTS}
Thanks to Derek Richardson and Mike A'Hearn for offering comments on the manuscript and to the anonymous referee for a careful and thorough review. 
MMK acknowledges support from various NASA grants. KJW acknowledges support from NLSI CLOE. 
Office space was generously provided for MMK by both the University of Maryland Department of Astronomy and Johns Hopkins University Applied Physics Laboratory while he conducted this work.


\begin{thebibliography}{}
\expandafter\ifx\csname natexlab\endcsname\relax\def\natexlab#1{#1}\fi

\bibitem[{{A'Hearn}(2011)}]{ahearn11b}
{A'Hearn}, M.~F. 2011, \araa, 49, 281

\bibitem[{{Asphaug} \& {Benz}(1996)}]{asphaug96}
{Asphaug}, E., \& {Benz}, W. 1996, Icarus, 121, 225

\bibitem[{{Biesecker} {et~al.}(2002){Biesecker}, {Lamy}, {St.~Cyr}, {Llebaria},
  \& {Howard}}]{biesecker02}
{Biesecker}, D.~A., {Lamy}, P., {St.~Cyr}, O.~C., {Llebaria}, A., \& {Howard},
  R.~A. 2002, Icarus, 157, 323

\bibitem[{{Boehnhardt}(2004)}]{boehnhardt04}
{Boehnhardt}, H. 2004, in Comets II, ed. M.~C. {Festou}, H.~U. {Keller}, \&
  H.~A. {Weaver} (Univ. of Arizona Press/Lunar Planet. Inst., Tucson,
  AZ/Houston, TX), 301--316

\bibitem[{{Boss} {et~al.}(1991){Boss}, {Cameron}, \& {Benz}}]{boss91}
{Boss}, A.~P., {Cameron}, A.~G.~W., \& {Benz}, W. 1991, \icarus, 92, 165

\bibitem[{{Bryans} \& {Pesnell}(2012)}]{bryans12}
{Bryans}, P., \& {Pesnell}, W.~D. 2012, \apj, 760, 18

\bibitem[{{Chandrasekhar}(1969)}]{chandrasekhar69}
{Chandrasekhar}, S. 1969, {Ellipsoidal figures of equilibrium} (Yale University
  Press, New Haven, CT)

\bibitem[{{Cowan} \& {A'Hearn}(1979)}]{cowan79}
{Cowan}, J.~J., \& {A'Hearn}, M.~F. 1979, Moon and Planets, 21, 155

\bibitem[{{Downs} {et~al.}(2013){Downs}, {Linker}, {Mikic}, {Riley},
  {Schrijver}, \& {Saint-Hilaire}}]{downs13}
{Downs}, C., {Linker}, J.~A., {Mikic}, Z., {et~al.} 2013, Science, 340, 1196

\bibitem[{{Gundlach} {et~al.}(2012){Gundlach}, {Blum}, {Skorov}, \&
  {Keller}}]{gundlach12}
{Gundlach}, B., {Blum}, J., {Skorov}, Y.~V., \& {Keller}, H.~U. 2012, ArXiv
  e-prints, arXiv:1203.1808

\bibitem[{{Holsapple} \& {Housen}(2007)}]{holsapple07}
{Holsapple}, K.~A., \& {Housen}, K.~R. 2007, \icarus, 187, 345

\bibitem[{{Holsapple} \& {Michel}(2008)}]{holsapple08}
{Holsapple}, K.~A., \& {Michel}, P. 2008, \icarus, 193, 283

\bibitem[{{Iseli} {et~al.}(2002){Iseli}, {K{\"u}ppers}, {Benz}, \&
  {Bochsler}}]{iseli02}
{Iseli}, M., {K{\"u}ppers}, M., {Benz}, W., \& {Bochsler}, P. 2002, Icarus,
  155, 350

\bibitem[{{Knight} {et~al.}(2010){Knight}, {A'Hearn}, {Biesecker}, {Faury},
  {Hamilton}, {Lamy}, \& {Llebaria}}]{knight10d}
{Knight}, M.~M., {A'Hearn}, M.~F., {Biesecker}, D.~A., {et~al.} 2010, \aj, 139,
  926

\bibitem[{{Kreutz}(1888)}]{kreutz88}
{Kreutz}, H. 1888, {Untersuchungen {\"U}ber das Cometensystem 1843 I, 1880 I
  und 1882 II.} (Kiel, Druck von C.~Schaidt, C.~F.~Mohr nachfl., 1888.), K92

\bibitem[{{Kronk}(2003)}]{kronk03}
{Kronk}, G.~W. 2003, {Cometography: A Catalog of Comets. Volume 2, 1800-1899}
  (Cambridge, UK: Cambridge University Press)

\bibitem[{{Kronk}(2009)}]{kronk09}
---. 2009, {Cometography: A Catalog of Comets. Volume 4, 1933-1959} (Cambridge,
  UK: Cambridge University Press)

\bibitem[{{Kronk} \& {Meyer}(2010)}]{kronk10}
{Kronk}, G.~W., \& {Meyer}, M. 2010, {Cometography: A Catalog of Comets. Volume
  5, 1960-1982} (Cambridge, UK: Cambridge University Press)

\bibitem[{{Lamy} {et~al.}(2004){Lamy}, {Toth}, {Fernandez}, \&
  {Weaver}}]{lamy04}
{Lamy}, P.~L., {Toth}, I., {Fernandez}, Y.~R., \& {Weaver}, H.~A. 2004, in
  Comets II, ed. M.~C. {Festou}, H.~U. {Keller}, \& H.~A. {Weaver} (Univ. of
  Arizona Press/Lunar Planet. Inst., Tucson, AZ/Houston, TX), 223--264

\bibitem[{{Li} {et~al.}(2013){Li}, {Weaver}, {Kelley}, {Farnham}, {A'Hearn},
  {Knight}, {Mutchler}, {Lamy}, \& {Toth}}]{cbet3496}
{Li}, J.-Y., {Weaver}, H.~A., {Kelley}, M.~S., {et~al.} 2013, Central Bureau
  Electronic Telegrams, 3496

\bibitem[{{Lisse} {et~al.}(2013){Lisse}, {Vervack}, {Weaver}, {Bauer},
  {Fernandez}, {Kelley}, {Knight}, {Hines}, {Li}, {Reach}, {Sitko},
  {Yanamandra-Fisher}, {Meech}, \& {Rayner}}]{cbet3598}
{Lisse}, C.~M., {Vervack}, R.~J., {Weaver}, H.~A., {et~al.} 2013, Central
  Bureau Electronic Telegrams, 3598

\bibitem[{{MacQueen} \& {St.~Cyr}(1991)}]{macqueen91}
{MacQueen}, R.~M., \& {St.~Cyr}, O.~C. 1991, Icarus, 90, 96

\bibitem[{{Marsden}(1967)}]{marsden67}
{Marsden}, B.~G. 1967, \aj, 72, 1170

\bibitem[{{Marsden}(1989)}]{marsden89}
---. 1989, \aj, 98, 2306

\bibitem[{{Marsden} \& {Williams}(2008)}]{marsden08}
{Marsden}, B.~G., \& {Williams}, G. 2008, {Catalogue of Cometary Orbits} (IAU
  Minor Planet Center/Central Bureau for Astronomical Telegrams)

\bibitem[{{McCauley} {et~al.}(2013){McCauley}, {Saar}, {Raymond}, {Ko}, \&
  {Saint-Hilaire}}]{mccauley13}
{McCauley}, P.~I., {Saar}, S.~H., {Raymond}, J.~C., {Ko}, Y.-K., \&
  {Saint-Hilaire}, P. 2013, \apj, 768, 161

\bibitem[{{Movshovitz} {et~al.}(2012){Movshovitz}, {Asphaug}, \&
  {Korycansky}}]{movshovitz12}
{Movshovitz}, N., {Asphaug}, E., \& {Korycansky}, D. 2012, \apj, 759, 93

\bibitem[{{Novski} \& {Novichonok}(2012)}]{cbet3238}
{Novski}, V., \& {Novichonok}, A. 2012, Central Bureau Electronic Telegrams,
  3238

\bibitem[{{Opik}(1966)}]{opik66}
{Opik}, E.~J. 1966, Irish Astronomical Journal, 7, 141

\bibitem[{{Raymond} {et~al.}(1998){Raymond}, {Fineschi}, {Smith}, {Gardner},
  {O'Neal}, {Ciaravella}, {Kohl}, {Marsden}, {Williams}, {Benna}, {Giordano},
  {Noci}, \& {Jewitt}}]{raymond98}
{Raymond}, J.~C., {Fineschi}, S., {Smith}, P.~L., {et~al.} 1998, \apj, 508, 410

\bibitem[{{Richardson} {et~al.}(1998){Richardson}, {Bottke}, \&
  {Love}}]{richardson98}
{Richardson}, D.~C., {Bottke}, W.~F., \& {Love}, S.~G. 1998, \icarus, 134, 47

\bibitem[{{Richardson} {et~al.}(2002){Richardson}, {Leinhardt}, {Melosh},
  {Bottke}, \& {Asphaug}}]{richardson02}
{Richardson}, D.~C., {Leinhardt}, Z.~M., {Melosh}, H.~J., {Bottke}, Jr., W.~F.,
  \& {Asphaug}, E. 2002, {Gravitational Aggregates: Evidence and Evolution},
  ed. {Bottke Jr., W.~F., Cellino, A., Paolicchi, P., \& Binzel, R.~P.} (Univ.
  of Arizona Press, Tucson, AZ), 501--515

\bibitem[{{Richardson} {et~al.}(2000){Richardson}, {Quinn}, {Stadel}, \&
  {Lake}}]{richardson00}
{Richardson}, D.~C., {Quinn}, T., {Stadel}, J., \& {Lake}, G. 2000, \icarus,
  143, 45

\bibitem[{{Richardson} {et~al.}(2007){Richardson}, {Melosh}, {Lisse}, \&
  {Carcich}}]{richardson07}
{Richardson}, J.~E., {Melosh}, H.~J., {Lisse}, C.~M., \& {Carcich}, B. 2007,
  Icarus, 190, 357

\bibitem[{{Roche}(1847)}]{roche47}
{Roche}, E.~A. 1847, Acad. des Sciences Lettres de Montpelier, 243

\bibitem[{{Samarasinha} \& {Mueller}(2013)}]{samarasinha13}
{Samarasinha}, N.~H., \& {Mueller}, B.~E.~A. 2013, ApJL, 775, L10

\bibitem[{{Samarasinha} {et~al.}(2004){Samarasinha}, {Mueller}, {Belton}, \&
  {Jorda}}]{samarasinha04}
{Samarasinha}, N.~H., {Mueller}, B.~E.~A., {Belton}, M.~J.~S., \& {Jorda}, L.
  2004, {Rotation of Cometary Nuclei}, ed. {Festou, M.~C., Keller, H.~U., \&
  Weaver, H.~A.} (Univ. of Arizona Press/Lunar Planet. Inst., Tucson,
  AZ/Houston, TX), 281--299

\bibitem[{{Schleicher}(2013{\natexlab{a}})}]{iauc9254}
{Schleicher}, D. 2013{\natexlab{a}}, \iaucirc, 9254

\bibitem[{{Schleicher}(2013{\natexlab{b}})}]{iauc9257}
---. 2013{\natexlab{b}}, \iaucirc, 9257

\bibitem[{{Schrijver} {et~al.}(2012){Schrijver}, {Brown}, {Battams},
  {Saint-Hilaire}, {Liu}, {Hudson}, \& {Pesnell}}]{schrijver12}
{Schrijver}, C.~J., {Brown}, J.~C., {Battams}, K., {et~al.} 2012, Science, 335,
  324

\bibitem[{{Sekanina}(1966)}]{sekanina66}
{Sekanina}, Z. 1966, Bulletin of the Astronomical Institutes of Czechoslovakia,
  17, 207

\bibitem[{{Sekanina}(2002)}]{sekanina02b}
---. 2002, \apj, 576, 1085

\bibitem[{{Sekanina}(2003)}]{sekanina03}
---. 2003, \apj, 597, 1237

\bibitem[{{Sekanina} \& {Chodas}(2007)}]{sekaninachodas07}
{Sekanina}, Z., \& {Chodas}, P.~W. 2007, \apj, 663, 657

\bibitem[{{Sekanina} \& {Chodas}(2012)}]{sekanina12}
---. 2012, \apj, 757, 127

\bibitem[{{Sridhar} \& {Tremaine}(1992)}]{sridhar92}
{Sridhar}, S., \& {Tremaine}, S. 1992, \icarus, 95, 86

\bibitem[{{Stern}(1990)}]{stern90}
{Stern}, S.~A. 1990, \icarus, 84, 447

\bibitem[{{Walsh} \& {Richardson}(2006)}]{walsh06}
{Walsh}, K.~J., \& {Richardson}, D.~C. 2006, \icarus, 180, 201

\bibitem[{{Weissman} {et~al.}(2004){Weissman}, {Asphaug}, \&
  {Lowry}}]{weissman04}
{Weissman}, P.~R., {Asphaug}, E., \& {Lowry}, S.~C. 2004, {Structure and
  density of cometary nuclei}, ed. {Festou, M.~C., Keller, H.~U., \& Weaver,
  H.~A.} (Univ. of Arizona Press/Lunar Planet. Inst., Tucson, AZ/Houston, TX),
  337--357

\end{thebibliography}

\label{lastpage}

\end{singlespace}

\renewcommand{\baselinestretch}{0.85}
\renewcommand{\arraystretch}{1.4}

\begin{deluxetable}{llll}  
\tabletypesize{\scriptsize}
\tablecolumns{4}
\tablewidth{0pt} 
\setlength{\tabcolsep}{0.05in}
\tablecaption{Factors affecting ISON's survivability}
\tablehead{   
  \multicolumn{1}{l}{Parameter}&
  \multicolumn{1}{l}{Value(s)}&
  \multicolumn{1}{l}{Qualitative Description of Results}&
  \multicolumn{1}{l}{Likely Value\tablenotemark{a}}
}
\startdata
Radius (km)&{\raise.3ex\hbox{$\scriptstyle<$}}0.2&Does not survive due to mass loss from sublimation&0.4--2.0\\
&0.2--1.0&Survival dependent on combination of other factors;\\
&&more likely to survive the larger the radius\\
&{\raise.3ex\hbox{$\scriptstyle>$}}1.0&Survives for most scenarios if density {\raise.3ex\hbox{$\scriptstyle>$}}0.1 g cm$^{-3}$\\
\hline

Density (g cm$^{-3}$)&{\raise.3ex\hbox{$\scriptstyle<$}}0.1&Does not survive for most scenarios&0.4--0.7\\
&0.1--0.7&Survival dependent on combination of other factors\\
&&more likely to survive the higher the density&\\
&{\raise.3ex\hbox{$\scriptstyle>$}}0.7&Survives for most scenarios\\
\hline

Axis ratio ($a:b$)&1.0&Survival dependent on combination of other factors;&1.0--2.6\\
&&more likely to survive the smaller the ratio $a:b$\\
&2.0&Survival dependent on combination of other factors\\
\hline

Sense of rotation&Prograde&Survival dependent on combination of other factors&Retrograde\\
&Retrograde&Survival for most scenarios&\\
&No spin&Survival dependent on combination of other factors;\\
&&more likely to survive than prograde case;\\
&&less likely to survive than  retrograde case\\
\hline

Rotation period (hr)&{\raise.3ex\hbox{$\scriptstyle>$}}P$_\mathrm{crit\_50\%}$&Survival dependent on combination of other factors;&4--180{\raise.3ex\hbox{$\scriptstyle+$}}\\
&&more likely to survive the longer the rotation period\\
&P$_\mathrm{crit\_50\%}$&Survival dependent on combination of other factors\\
&{\raise.3ex\hbox{$\scriptstyle<$}}P$_\mathrm{crit\_50\%}$&Survival dependent on combination of other factors\\
\hline
\enddata
\tablenotetext{a} {See the text for the sources of these values.}
\label{t:parameters}
\label{lasttable}
\end{deluxetable}


\end{document}